# Construction of Chiral Metamaterial with a Helix Array

Xiang Xiong[1], Xiao-Chun Chen[1], Zhao-Wu Wang[1], Shang-Chi Jiang[1], Mu Wang[1,*],
Ru-Wen Peng[1], Xi-Ping Hao[1], and Cheng Sun[2]

[1]*National Laboratory of Solid State Microstructures and Department of Physics,
Nanjing University, Nanjing 210093, China*
[2]*Department of Mechanical Engineering, Northwestern University, Evanston, IL 60208-3111, USA*

Here we report the designing of chiral metamaterial with metallic helix array. The effective electric and magnetic dipoles, which originate from the induced surface electric current upon illumination of incident light, are collinear at the resonant frequency. Consequently, for the circularly polarized incident light, negative refractive index is realized. Our design provides a unique approach to tune the optical properties by assembling helices, and demonstrates a different approach in exploring three-dimensional chiral metamaterial.

PACS numbers:   78.20.Ek, 78.67.-n, 42.25.Ja, 73.20.Mf

Keywords:   metamaterial, helix, chiral, negative refractive index

The interest to construct metamaterial has being promoted in recent years by its attractive optical properties, such as negative refractive index[1-4], ultrahigh spatial resolution[5-7], invisiblity cloaking[8-10], and negative optical pressure[11], etc. One of the interesting structures is chiral metamaterial[12-16], which offers a unique way to realize negative refractive index. In chiral materials, the structural chirality helps to suppress the refractive index of light with one handedness, and increases the refractive index of light with the opposite handedness. When chirality of the structure becomes sufficiently strong, negative refractive index can be realized[12]. So far, a number of chiral structures, such as cross-wires[17], twisted rosettes[15], interlocked spilt-ring resonators[16] and U-shaped resonators[13], etc., have been proposed to construct the negative-refractive-index metamaterials.

In chiral metamaterial the strongly coupled electric and magnetic dipoles are simultaneously excited. In previous designings, there usually exists an angle between the electric and magnetic dipoles[14], which makes it difficult to utilize the induced dipoles efficiently. We once proposed an assembly of double-layered metallic U-shaped resonators[13], which possesses two resonant frequencies $\omega_H$ and $\omega_L$, respectively. The effective electric and magnetic dipoles, which originate from the specific distrubution of induced surface electric current upon illumination, are colinear at the resonant frequency. Consequently, for the left-handed circularly polarized incident light, negative refractive index could occurs at $\omega_H$, whereas for the right-handed circularly polarized incident light, negative refractive index occures at $\omega_L$. Despite of efficient utilization of the electric and magnetic dipoles in that approach, the coupling between differernt layers of the buliding block is very strong, which is not a favorable feature for constructing three-dimensional (3D) metamaterial. One possible solution to solve this problem is to introduce helix structure[18,19], where collinear electric and magnetic dipoles can be parallelly or anti-parallelly excited. The coupling between different layers of helix structure is small. Therefore, by combining helices in a specific way, 3D chiral metamaterials can be constructed.

Here we show that in an assembly of helix array, the effective induced electric and magnetic dipoles are aligned. The resulted strong chiral feature leads to the negative refraction for circularly polarized light. The elementary buliding block is a uniaxial gold helix with three and a half turns, as shown in Fig. 1(a). The axis of the helix is along *x*-direction and the two endpoints of the helix are positioned in *x-z* plane. The wave vector of the incident light is along *z*-direction. Commercial software based on the finite difference time domain (FDTD) method (CST Microwave Studio) is applied in calculation. The permittivity of gold in the infrared regime is based on the Drude model, $\varepsilon(\omega)=1-\omega_p^2/(\omega^2+i\omega_\tau\omega)$, where $\omega_p$ is the plasma frequency, and $\omega_\tau$ is the damping constant. These parameters are taken as $\omega_p=1.37\times10^4$ THz, and $\omega_\tau=40.84$ THz[20]. Considering that the electrons may experience additional scattering from metal surfaces, in the simulation we take the damping constant 1.6 times higher than that of the bulk material[21]. Figure 1(b) shows the transmission coefficients of the array of helix unit shown in Fig. 1(a). The distance between neighboring helices in both *x* and *y* directions is 0.05 $\mu$m. The resonant dips in transmission $t_{//}$, where the polarization of input and output waves is in parallel, can be detected at 1300 cm$^{-1}$ for both *x*-polarized (**t**$_{x/\!/}$) and *y*-polarized (**t**$_{y/\!/}$) incidence. It should be noted that the helix possesses intrinsic chirality and the helix array shows optical activity. The chiral behavior of the helix array rotates the polarization of incident light and converts a portion of energy from one polarization to the other. Consequently, the resonance peaks of perpendicular polarization of input and output light (**t**$_\perp$) can be detected in the transmission at 1300 cm$^{-1}$ for both *x* polarization (**t**$_{x\perp}$) and *y* polarization (**t**$_{y\perp}$). Calculations show that **t**$_{x\perp}$ and **t**$_{y\perp}$ are almost identical. There may exist higher order of resonances in the helix structure, and the induced surface electric current forms a much more complicated pattern (i.e., the current flows in different directions on different section of the helix). Here we focus, however, on the lowest frequency resonance of the structure (1300 cm$^{-1}$), where the induced surface electric current flows in the same direction, as shown in Fig. 1(c). The resonant mode of the surface electric current in the helix does not depend on polarization direction. The surface current shown in Fig. 1(c) essentially flows along *x*-direction, which corresponds to an effective electric field $E'$ in *x*, as indicated by red (light gray) arrow in Fig. 1(d). This means that an effective electric dipole along *x* is induced. On the other hand, the surface electric current along helix forms a loop structure. The curl integration along the projection in *y-z* plane is nonzero, which leads to an induced magnetic field $H'$ in *x*. This indicates that an effective magnetic dipole along *x* is induced at the same time, as indicated by the blue (dark gray) arrow in Fig. 1(d).

Four helices are assembled into a unit with each helix being seperated 0.05 $\mu$m apart, as shown in Fig. 2(a). An array of such units arranged in simple square lattice is then constructed, and the lattice parameter equals to 0.7 $\mu$m in both *x*- and *y*-directions. Due to the fourfold rotatioanl symmetry of the unit, the transmisssion and reflection properties of this structure do not rely on the orientation of the sample with respect to the polarization of incident light. The strucutre shown in Fig. 2(a) is clearly different from what we reported earlier[18]. For previous structure in ref. 18, the magentic and electric responses can be switched at the same frequency by changing the polarization of the incident light. For the chiral structure shown in Fig. 2(a), software is applied to calculate the transmission and reflection coefficients, and the results are shown in Fig. 2(b). It follows that one resonance dip occurs at 1300 cm$^{-1}$, which corresponds to the transmission of parallel polarization of input and out put light ($|t_{//}|$) and to the transmission of perpendicular polarization of input and output light ($|t_{\perp}|$). For optically inactive structure, when the incident light is polarized along the principal axis (the axis along which the tensor of electric susceptibility is diagonal), only transmission of $t_{//}$ is detected while $t_{\perp}$ vanishes. For optically active material as we discuss here, however, the chiral behavior of the structure rotates the polarization of incident light and convert a portion of energy from one polarization to the other polarization. Hence $t_{\perp}$ can be detected, as illustrated in Fig.2 (b).

Let us define the transmission and reflection coefficients of the left-handed circularly polarized (LCP) light and the right-handed circularly polarized (RCP) light as $t_L=t_{//}-it_{\perp}$, $t_R=t_{//}+it_{\perp}$, $r_L=r_{//}-ir_{\perp}$, $r_R=r_{//}+ir_{\perp}$, respectively. In our structure $r_{\perp}$ is zero, as shown in Fig. 2(b), suggesting that the polarization of the reflection light does not change.

The calculated transmission coefficients of LCP and RCP ($t_L$ and $t_R$) are shown in Fig. 2(c). Due to the chirality of the structure, the transmission for LCP and RCP split into two different curves (Fig. 2(c)). One resonant dips appears at 1300 cm$^{-1}$ for both $t_L$ and $t_R$, respectively. For the resonance at 1300 cm$^{-1}$, the dip in $|t_L|$ is much deeper than that in $|t_R|$, indicating that the resonance for LCP is stronger than that for RCP. We define here $\delta$ as the phase difference between $t_\perp$ and $t_{//}$. The dependence of $\delta$ as a function of wave number is illustrated in Fig. 2(d). It follows that an elliptical polarized light is generated when $\delta \neq n\pi$ (n is an integer). Meanwhile, the end of electric field vector revolves clockwise for $\sin\delta >0$ and count-clockwise for $\sin\delta <0$. Consequently, the outgoing light revolves clockwise in our design. The azimuth angle of the principal polarization axis of the transmitted wave, $\theta$, is defined as $\theta = \frac{1}{2}[\arg(t_R)-\arg(t_L)]$, which represents the change of polarization angle when a linearly polarized light is incident on the helices assembly[22]. As plotted in Fig. 2(d), at 1350 cm$^{-1}$ the azimuth angle reaches 22°.

It is known that the refractive indices for LCP and RCP lights are given by $n_{R/L} = \sqrt{\varepsilon\mu} \pm \xi$ [14,15], where $\varepsilon$ and $\mu$ are the effective permittivity and permeability, respectively. The chiral parameter $\xi$, which is defined as $\xi =(n_R-n_L)/2$, represents the coupling between the electric and the magnetic dipoles along the same direction. The effective impedance (Z) and refractive index for LCP and RCP light can be derived from the reflection and transmission coefficient[15,23]. It is noteworthy that in Fig. 3(a)-(b), for our helix structure, evident drop of refractive index occurs at 1300 cm$^{-1}$ for LCP light and the refractive index shows much smaller modification for RCP light.

This is due to the cancellation of the contributions from the permittivity/permeability and that from the chirality. Far away from the resonant frequency, the difference between the refractive indices of RCP and LCP diminishes. The effective permittivity and permeability have been retrieved, as illustrated in Fig. 3(c)-(d). At 1300 cm$^{-1}$ the permittivity reaches negative values, whereas the magnetic resonance is not sufficiently strong to provide a negative permeability. The chiral term $\xi$ and impedance Z are shown in Fig. 3(e)-(f), respectively, which demonstrate strong chirality of our helix array at 1300 cm$^{-1}$.

In U-shaped resonator (USR) structure in our previous studies[13, 25], some surface current loops are induced in *x-y* plane along USR arms, strong magnetic dipoles are induced in *z*-direction. When different layers of USR are stacked along *z*-direction, strong coupling of different layers prevents homogenization of the metamaterial[26]. For helix structure, the induced electric and magnetic dipoles are almost along *x* or *–x*. The electromagnetic filed along *z*-direction is relatively small, which means the coupling in *z*-direction is weak. There is no strong magnetic field of one helix penetrating into the neighboring one. So we are able to stack layers of helix array to generate 3D metamaterial. Figure 4(a) shows the unit cell of a double-layer helix array. To construct the 3D structure, we duplicate the first layer of helix array, and then rotate each helix for 90° counterclockwise around *z*-axis. Then we lift the duplicated helix array upwards by 0.5 $\mu$m to form the second layer, as shown in Fig. 4(a). The third layer can be produced in the same way. The azimuth angle of the principle axis of polarization of transmission wave, $\theta$, as a function of wave number for single-,

double- and triple-layered structure is shown in Fig. 4 (b), which shows the change of polarization angle when a linearly polarized light is incident on the helix assembly. Around resonant frequency, the multi-layered chiral structure shows stronger optical activity than the single-layered one. As shown in Fig. 4(b), at resonant frequency the rotation angle reaches 42° for the double-layer helix array and 60° for triple-layer helix array. The chiral parameter $\xi$ and the refractive index for single-, double- and triple-layered structure are shown in Fig. 4(c)-(d), respectively. It can be seen that negative refractive index can be realized for LCP light for the 3D helix array.

This work was supported by grants from the NSF of China (Grant Nos. 50972057, 10874068, 11034005, and 61077023), the State Key Program for Basic Research from MOST of China (Grant No. 2010CB630705), and partly by Jiangsu Province (Grant No. BK2008012).

[1] R. A. Shelby, D. R. Smith, and S. Schultz, Science **292**, 77 (2001).

[2] S. Zhang, W. J. Fan, N. C. Panoiu, K. J. Malloy, R. M. Osgood, and S. R. J. Brueck, Phys. Rev. Lett. **95**, 137404 (2005).

[3] V. M. Shalaev, Nat. Photonics **1**, 41 (2007).

[4] J. Valentine, S. Zhang, T. Zentgraf, E. Ulin-Avila, D. A. Genov, G. Bartal, and X. Zhang, Nature **455**, 376 (2008).

[5] J. B. Pendry, Phys. Rev. Lett. **85**, 3966 (2000).

[6] N. Fang, H. Lee, C. Sun, and X. Zhang, Science **308**, 534 (2005).

[7] N. Fang and X. Zhang, Appl. Phys. Lett. **82**, 161 (2003).

[8] J. B. Pendry, D. Schurig, and D. R. Smith, Science **312**, 1780 (2006).

[9] R. Liu, C. Ji, J. J. Mock, J. Y. Chin, T. J. Cui, and D. R. Smith, Science **323**, 366 (2009).

[10] T. Ergin, N. Stenger, P. Brenner, J. B. Pendry, and M. Wegener, Science **328**, 337 (2010).

[11] H. Liu, J. Ng, S. B. Wang, Z. F. Lin, Z. H. Hang, C. T. Chan, and S. N. Zhu, Phys. Rev. Lett. **106**, 087401 (2011).

[12] J. B. Pendry, Science **306**, 1353 (2004).

[13] X. Xiong, W. H. Sun, Y. J. Bao, M. Wang, R. W. Peng, C. Sun, X. Lu, J. Shao, Z. F. Li, and N. B. Ming, Phys. Rev. B **81**, 075119 (2010).

[14] S. Zhang, Y. S. Park, J. S. Li, X. C. Lu, W. L. Zhang, and X. Zhang, Phys. Rev. Lett. **102**, 023901 (2009).

[15] E. Plum, J. Zhou, J. Dong, V. A. Fedotov, T. Koschny, C. M. Soukoulis, and N. I.

Zheludev, Phys. Rev. B **79**, 035407 (2009).

[16] B. N. Wang, J. F. Zhou, T. Koschny, and C. M. Soukoulis, Appl. Phys. Lett. **94**, 151112 (2009).

[17] J. F. Zhou, J. F. Dong, B. N. Wang, T. Koschny, M. Kafesaki, and C. M. Soukoulis, Phys. Rev. B **79**, 121104 (2009).

[18] X. Xiong, X. C. Chen, M. Wang, R. W. Peng, D. J. Shu, and C. Sun, Appl. Phys. Lett. **98**, 071901 (2011).

[19] J. K. Gansel, M. Thiel, M. S. Rill, M. Decker, K. Bade, V. Saile, G. von Freymann, S. Linden, and M. Wegener, Science **325**, 1513 (2009).

[20] M. A. Ordal, R. J. Bell, R. W. Alexander, L. L. Long, and M. R. Querry, Appl. Opt. **24**, 4493 (1985).

[21] S. Linden, C. Enkrich, M. Wegener, J. F. Zhou, T. Koschny, and C. M. Soukoulis, Science **306**, 1351 (2004).

[22] J. D. Jackson, *Classical Electrodynamics* (Wiley, New York, 1999).

[23] D. R. Smith, S. Schultz, P. Markos, and C. M. Soukoulis, Phys. Rev. B **65**, 195104 (2002).

[24] S. M. Xiao, V. P. Drachev, A. V. Kildishev, X. J. Ni, U. K. Chettiar, H. K. Yuan, and V. M. Shalaev, Nature **466**, 735 (2010).

[25] X. Xiong, W. H. Sun, Y. J. Bao, M. Wang, R. W. Peng, C. Sun, X. Lu, J. Shao, Z. F. Li, and N. B. Ming, Phys. Rev. B **80**, 201105(R) (2009).

[26] A. Andryieuski, C. Menzel, C. Rockstuhl, R. Malureanu, F. Lederer, and A. Lavrinenko, Phys. Rev. B **82**, 235107 (2010).

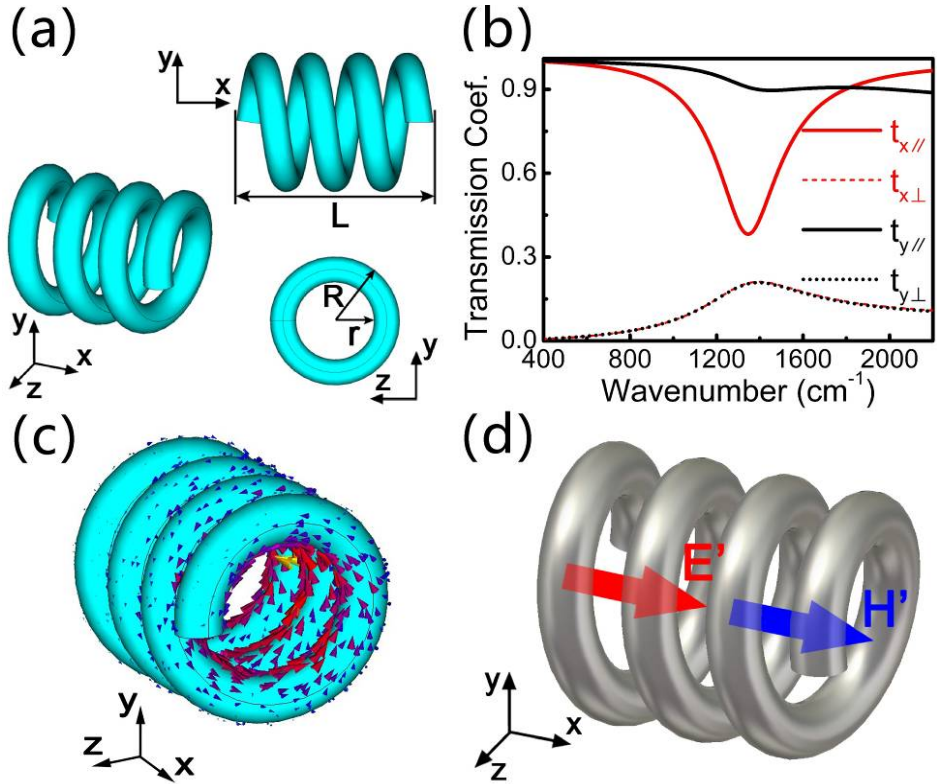

FIG. 1. (Color online) (a) The geometrical parameters of building block: L=0.35 $\mu$m, R=0.125 $\mu$m, and r= 0.075 $\mu$m. (b) Transmission coefficients of an array of helices for *x*- and *y*-polarized incidence. (c) shows the calculated induced surface electric current density on the helix. (d) schematically show the equivalent electric and magnetic dipoles induced on the helix.

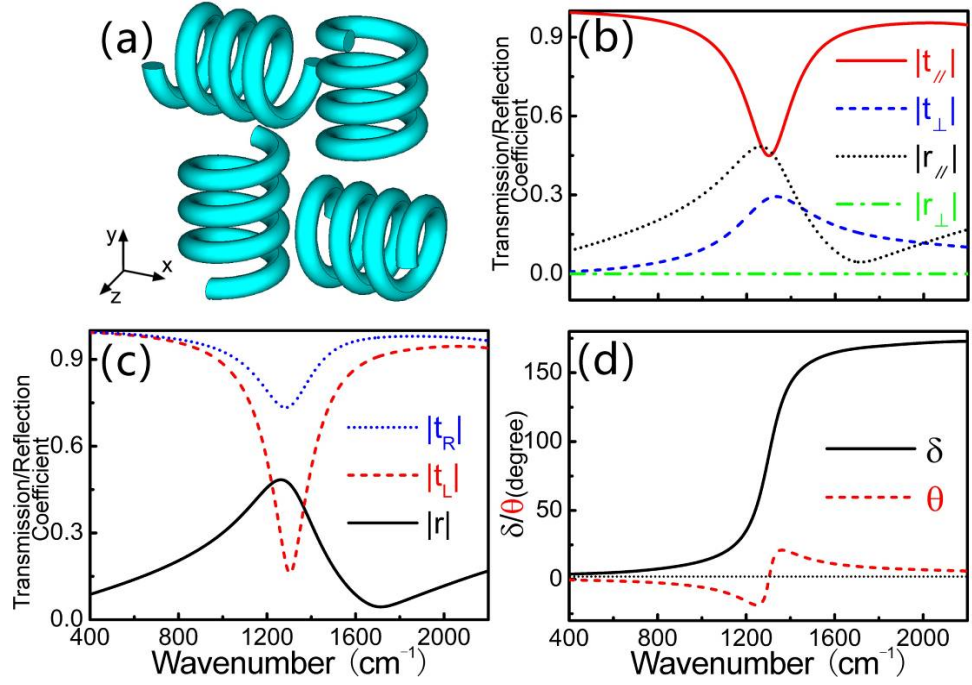

FIG. 2. (Color online) (a) The unit cell constructed by four helices. (b) The amplitudes of $t_\parallel, t_\perp, r_\parallel$ and $r_\perp$. (c) The amplitude of $t_L$, $t_R$ and $r$. (d) The diagram to show the phase difference $\delta$ between $t_\perp$, $t_\parallel$ as a function of wave number, and the azimuth angle of the principle axis of polarization of transmission wave $\theta$ as a function of wave number.

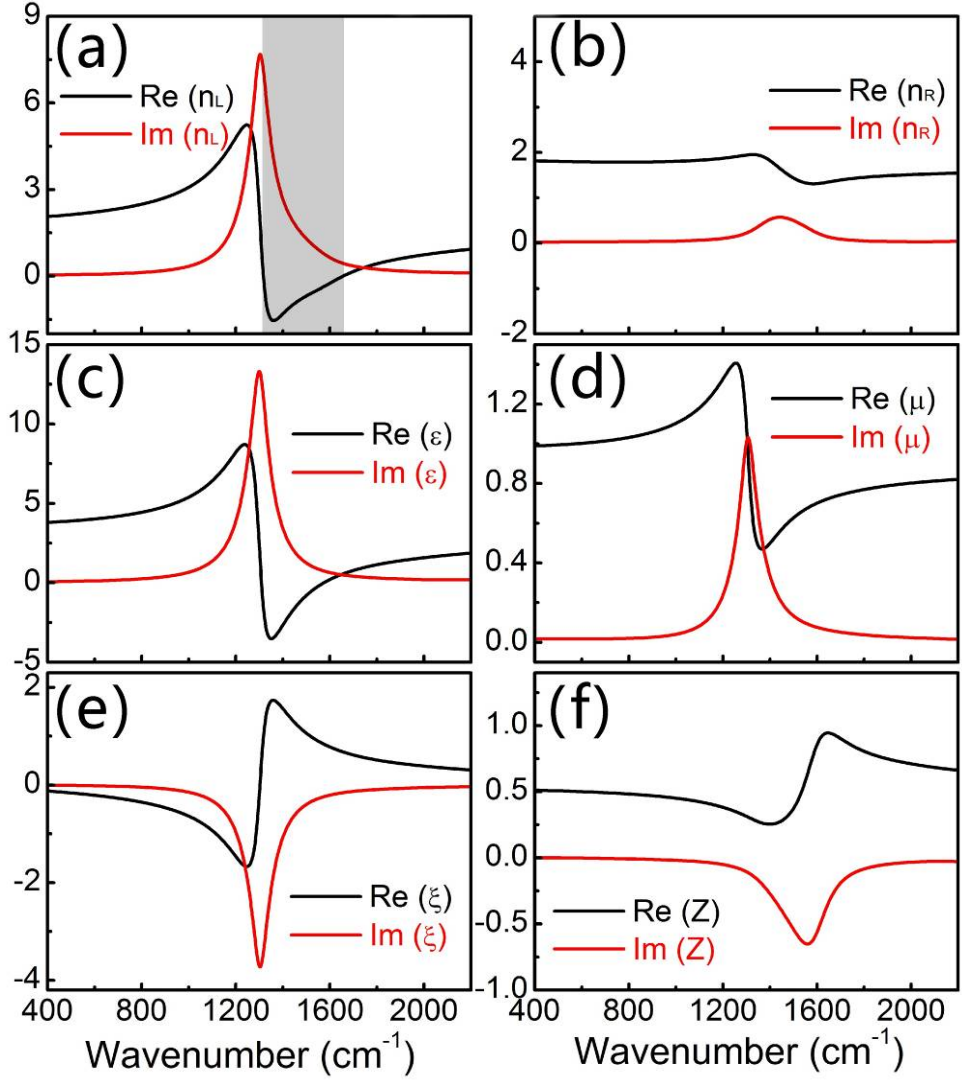

FIG. 3. (Color online) The retrieved effective optical parameters of helix arrays. (a) and (b) illustrate the real and imaginary parts of the refractive index for LCP and RCP light, respectively. The shadow denotes the occurrence of negative refractive index. (c)-(f) illustrate the real and imaginary parts of the permittivity $\varepsilon$, permeability $\mu$, chiral parameter $\xi$, and the impedence $Z$.

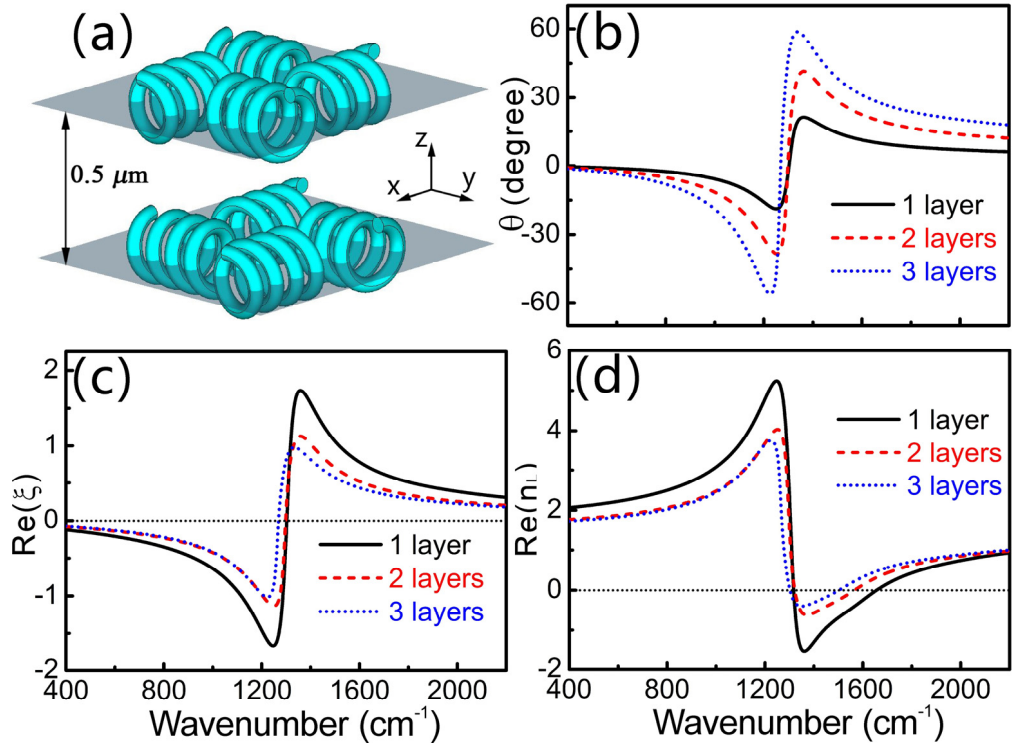

FIG. 4. (Color online) (a) The unit cell of the double-layer helix array. (b) the azimuth angle $\theta$ as a function of wave number for single-, double- and triple- layer helix arrays. (c) The real parts of chiral parameters and (d) the real parts of the refractive index for LCP for single-, double- and triple- layer helix arrays, separately.